\documentclass[%
 reprint,
superscriptaddress,
 amsmath,amssymb,
 aps,
]{revtex4-2}

\usepackage{graphicx}% Include figure files
\usepackage{dcolumn}% Align table columns on decimal point
\usepackage{bm}% bold math
\begin{document}

\preprint{APS/123-QED}

\title{On-chip time-domain terahertz spectroscopy of superconducting films below the diffraction limit}

\author{Alex M. Potts}
\altaffiliation{These authors contributed equally to this work}
\affiliation{Department of Physics, University of California at Santa Barbara, Santa Barbara CA 93106, USA}

\author{Abhay K. Nayak}%
\altaffiliation{These authors contributed equally to this work}
\affiliation{Department of Physics, University of California at Santa Barbara, Santa Barbara CA 93106, USA}%

\author{Michael Nagel}
\affiliation{Protemics GmbH, 52074 Aachen, Germany}

\author{Kelson Kaj}
\affiliation{Department of Physics, University of California at San Diego, La Jolla, CA 92093, USA}%

\author{Biljana Stamenic}
\affiliation{Nanofabrication Facility, Department of Electrical and Computer Engineering, University of California, Santa Barbara, California 93106, USA}%

\author{Demis D. John}
\affiliation{Nanofabrication Facility, Department of Electrical and Computer Engineering, University of California, Santa Barbara, California 93106, USA}%

\author{Richard D. Averitt}
\affiliation{Department of Physics, University of California at San Diego, La Jolla, CA 92093, USA}%

\author{Andrea F. Young}
\email{andrea@physics.ucsb.edu}
\affiliation{Department of Physics, University of California at Santa Barbara, Santa Barbara CA 93106, USA}

\date{\today}% It is always \today, today,

\begin{abstract}
  Free-space time domain THz spectroscopy accesses electrodynamic responses in a frequency regime ideally matched to interacting condensed matter systems. However, THz spectroscopy is challenging when samples are physically smaller than the diffraction limit of $\sim$0.5 mm, as is typical, for example, in van der Waals materials and heterostructures. Here, we present an on-chip, time-domain THz spectrometer based on semiconducting photoconductive switches with a bandwidth of 200 GHz to 750 GHz. We measure the optical conductivity of a 7.5-um wide NbN film across the superconducting transition, demonstrating spectroscopic signatures of the superconducting gap in a sample smaller than 2\% of the Rayleigh diffraction limit.  Our spectrometer features an interchangeable sample architecture, making it ideal for probing superconductivity, magnetism, and charge order in strongly correlated van der Waals materials. 

\end{abstract}

\maketitle

Terahertz time-domain spectroscopy (THz TDS) can be used to extract the complex optical conductivity of quantum materials without Kramers-Kronig relations. The few-meV energy scale and the picosecond time scale of a typical THz pulse makes THz spectroscopy well suited to experiments in many fields of research, particularly strongly correlated phenomena \cite{Basov2011, Valdes2012, Wu2016, Kastl2015}, superconductivity \cite{Bilbro2011, Fausti2011, Hu2014, Mahmood2019} and non-equilibrium dynamics\cite{Fausti2011, Mitrano2016}. 
However, free space THz TDS, which has been used extensively to study bulk materials and thin films, cannot be used directly to study correlated phenomena in van der Waals heterostructures, an increasingly active area for probing these same phenomena \cite{Kennes2021, Balents2020, Andrei2021}. Free space THz TDS inherently suffers from the inability to focus THz beams on samples smaller than the Rayleigh diffraction limit of 488 $\mu$m at 0.75 THz. vdW heterostructures, often $10$ $\mu$m or smaller, have a reduced cross-section with free-space THz beams, leading to a drastically diminished signal-to-noise ratio that precludes most experiments. Though THz near-field imaging techniques \cite{Schaffer2021, Eisele2014,Cocker2013, Jelic2017, Mittleman2018, Fei2011} can achieve extreme sub-wavelength resolution, they are often difficult to integrate into cryostats, have low excitation efficiency if standard far-field excitation techniques are used, and are limited to narrow bands of THz frequencies.

The diffraction limit can also be circumvented by confining THz pulses within transmission lines patterned on a chip. Early on-chip THz experiments incorporated photoconductive (PC) switches into transmission line circuits to emit and detect THz radiation\cite{Auston1975, Auston1988, Grischkowsky2000}. Recent on-chip THz efforts\cite{Wood2013, Cunningham2010, Dazhang2009, Heligman2021, Karnetzky2018, Shi2022, Smith2019, Wood2006, Yoshioka2022, Wang2023} have explored new geometries, leading to increases in the THz amplitude and spectral bandwidth. On-chip THz TDS has been recently used to study electronic phases at cryogenic temperatures in carbon nanotubes \cite{Zhong2008}, topological insulators \cite{Kastl2015}, and two-dimensional electronic systems such as GaAs \cite{Wood2012} and graphene \cite{McIver2019, Gallagher2019, Island2020, Hunter2015, Prechtel2012}. Most of the existing designs for on-chip THz spectroscopy require monolithic fabrication of the THz emitter/receiver and the vdW device. The combined complexity of fabricating both the THz spectrometer and a vdW device to perform a single experiment, however, reduces measurement throughput.

\begin{figure*}
\includegraphics[width=\textwidth]{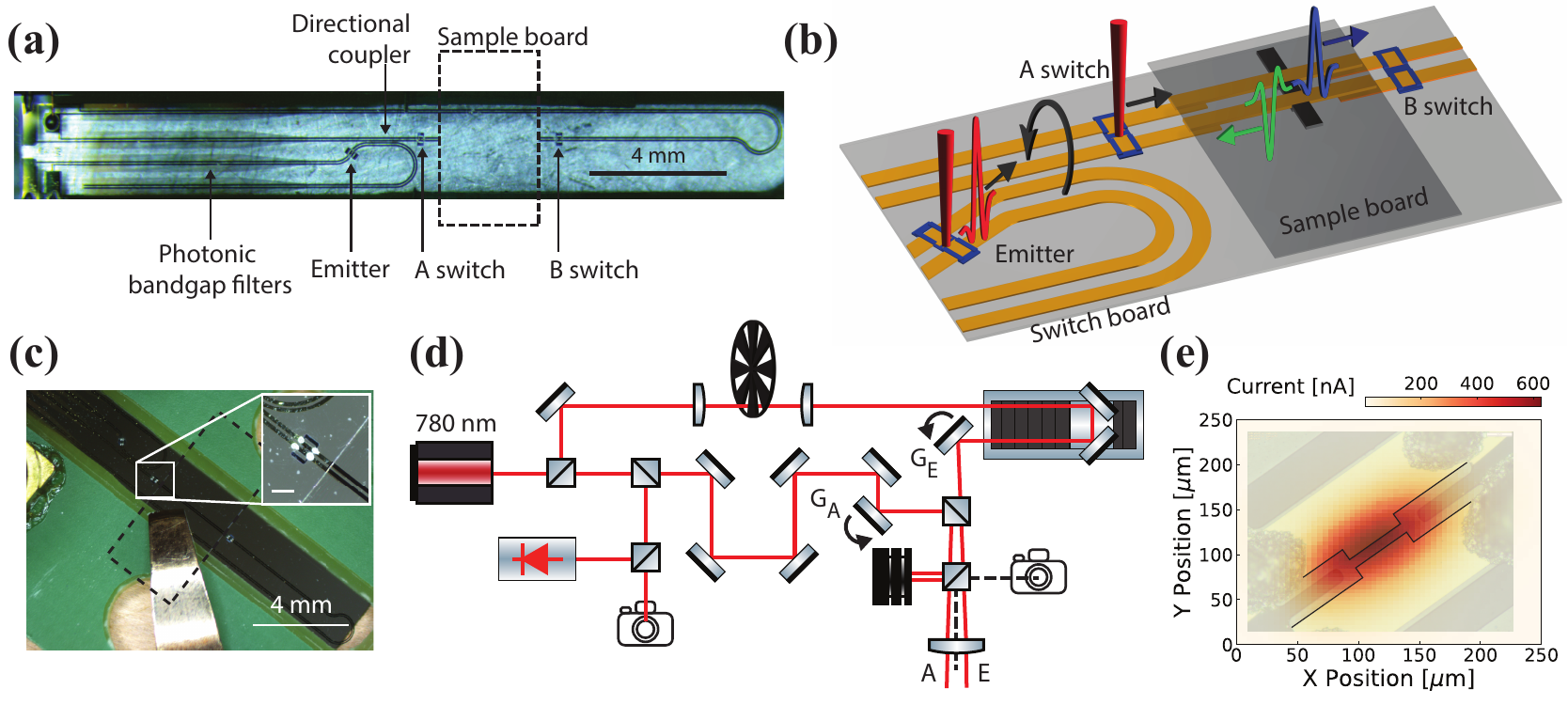}
\caption{On-chip THz spectrometer with fast sample exchange architecture. (a) Micrograph of the switch board, showing key circuit components and three LT GaAs PC switches (emitter, A-switch, and B-switch) with the location of the replaceable sample board indicated. (b) Schematic illustration of the on-chip THz spectrometer, the switch-sample board architecture, the emitted propagating transient (red), reflected transient (green) and transmitted transient (blue). (c) Micrograph of a thru sample board (outlined by the black dashed box) mounted on the switch board. The sample board transmission lines are precisely aligned to the switch board transmission lines, as shown in the inset (scale bar: 250 $\mu$m). A metal clip holds the sample board firmly in place to ensure excellent coupling of the switch board and sample board transmission lines. (d) Block optics diagram, showing a pulsed laser split into two paths. One beam is chopped, delayed, and directed by scanning galvanometer G$_E$ to the emitter (E) switch. The other beam is statically delayed and steered by scanning galvanometer G$_A$ to the readout (A) switch. The reflection from the sample is directed to a camera through the beamsplitter, used for coarse alignment. Some power from the A-switch beam is siphoned for power and Poynting stability monitoring. (e) Dynamic alignment scan map, showing recorded DC current versus galvanometer-controlled beam position with overlaid micrograph of the emitter photocunductive switch. Black lines denote the edges of gold transmission lines.}
\label{fig:architecture}
\end{figure*}

Higher throughput may be achieved by disentangling the fabrication of the THz spectrometer (switch board) and the device (on the sample board) as introduced recently by Lee \textit{et al.} \cite{Lee2022} at room temperature. Here, we demonstrate cryogenic operation of an on-chip THz spectrometer with a fast sample exchange architecture, and use it to extract the complex optical conductivity of a lithographically-defined NbN device below its superconducting transition temperature. Our experimental geometry is shown in Fig.\ref{fig:architecture}, and consists of a `switch board' containing transmission lines and photoconductive switches for both THz generation and readout and an interchangeable `sample board' containing a sample of interest, transmission line segments, and additional deposited electrical contacts.

The switch board is a permanent fixture containing low temperature (LT) grown GaAs PC switches coupled to lithographically defined coplanar stripline transmission lines (Fig.\ref{fig:architecture}a). The transmission lines have a conductor width of $50$ $\mu$m, a separation of $30$ $\mu$m, and are terminated by photonic bandgap filters that reduce THz reflections from the transmission line terminations at long times. The switch board is fabricated on a $50$ $\mu$m thick cyclo-olefin polymer (COP) substrate. High transparency, low permeability and a nearly temperature and frequency-independent optical index of $\approx$1.52 at THz frequencies ensures minimal pulse broadening \cite{Yanagi2008, Cunningham2011, Kaji2018}. The thin substrate and comparatively small dielectric constant substrate mitigates the $f^3$-dependent Cerenkov radiation loss that often dominates on-chip THz propagation \cite{Auston1983, Grischkowsky1987, Sychugin2019, Wang2015, Smith2019, Grischkowsky2000}. In addition, the local dielectric environment of thin COP substrates does not support unbound leaky-wave modes (see supplementary).

%leaky wave IS a technical term, no need for quotation marks

We use separate transmission lines for THz emission and detection. The transmission line for the THz detection is gapped between the detector A-switch and B-switch to allow for the placement of the sample board. Sample boards are fabricated on 50 $\mu$m thick z-cut dual-side polished quartz. Nanofabrication of fragile 50 $\mu$m quartz is accomplished by crystal-bonding quartz to a silicon carrier wafer, patterning the coplanar stripline (CPS) and performing any additional sample-related nanofabrication steps, and then dissolving the crystal bond to release the quartz film. 
The sample board is then aligned to the switch board under an optical microscope, resulting in a reliable placement precision of $\leq$10 $\mu$m. The resulting combined structure is shown schematically in Fig.\ref{fig:architecture}b while a micrograph of an aligned thru sample board is shown in Fig.\ref{fig:architecture}c.

We illuminate both the emitter and detector (A/B) PC switches with 3 mW of 780 nm, 75 fs Gaussian laser pulses with a repetition rate of 250 MHz. The optical path is illustrated in Fig.\ref{fig:architecture}d. The beam to the emitter switch is chopped at $\sim$3.2 kHz, routed through an optical delay stage, directed on the emitter switch using a scanning galvanometer (G$_E$), and focused to diameter $\approx25$ $\mu m$. A beam-splitter  extracts $\sim$500 $\mu$W for power and spot stability monitoring on the A-switch beam path. We apply a 10V DC bias across the emitter switch, providing the energy that is converted into a THz transient in the emission line when the emitter switch is illuminated. This THz pulse is subsequently coupled to the detection transmission line via an inductive directional coupler, which provides a DC block between the emission and detection transmission lines (Fig.\ref{fig:architecture}b). To improve the stability of our measurements, we spatially map the DC photocurrent across the emitter switch using scanning galvanometer G$_E$ (Fig.\ref{fig:architecture}e), allowing us to stabilize switch operation at the position of maximum photocurrent through real-time feedback. 

The beam for the detector switch is routed through another scanning galvanometer (G$_A$) prior to being focused on the switch, thus allowing for automatic and precise alignment of the laser beam on the PC switches. The THz pulse traveling along the detection transmission line is then measured at the detector A-switch. The detector switch is momentarily triggered by an optical pulse and thus allows us to measure the current across the detector switch, which samples the local electric field across the detection PC switch. Since the instantaneous current across the detector A-switch is only hundreds of picoamps, we use a transimpedance amplifier with gain 1e-7 A/V, the output of which is then demodulated using a lock-in amplifier synchronized to the chopper. Sweeping the optical delay stage continuously allows us to vary the time delay of the optical pulses exciting the emitter and the detector switches. This allows us to extract the time-domain electric field of the incident, reflected, and transmitted THz pulse from the transients recorded at the PC switch A and B. We move the optical delay stage continuously, measuring the time-domain THz signal with a Nyquist frequency of nearly 350 THz. We, then, apply standard THz TDS signal processing \cite{Neu2018, Potts2019} and smooth the time-domain data with a 101-point, 3rd order Savitsky-Golay filter.

\begin{figure*}
\includegraphics[width = \linewidth]{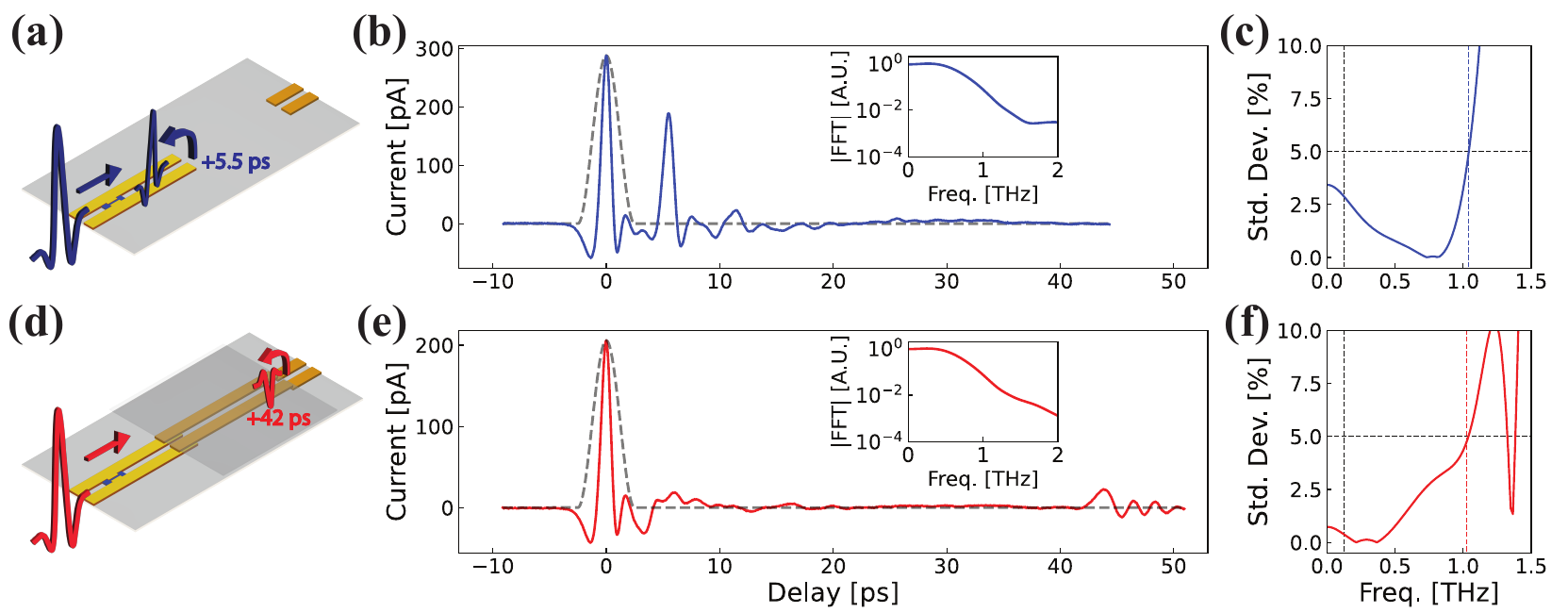}
\caption{THz reflection transients for two simple samples at room temperature. (a) Schematic showing incident transient on bare switch board and reflection from open circuit termination. (b) Time domain electric field with no sample board, showing the directivity pulse, corresponding isolating window (in dashed lines), and delayed reflections corresponding to the switch-sample board interface and a round trip from the second switch-sample board interface (inset: Fourier amplitude of system directivity $D(\omega)$). (c) 5\% magnitude repeatability, showing a bandwidth of 1.05 THz. (d) Schematic showing incident transient with a mounted thru sample board. (e) Time domain electric field with thru sample board, showing the directivity pulse and corresponding window (in dashed lines), a minimal first switch-sample board interface reflection, and a delayed reflection from the second switch-sample board interface (inset: Fourier amplitude of system directivity). (f) 5\% magnitude repeatability, showing a bandwidth of 1 THz.}
\label{fig:signals}
\end{figure*}

Fig. \ref{fig:signals}a-c show characterization data of the spectrometer at room temperature with no sample board mounted. In this configuration, the THz pulse traverses the A-switch (producing a `directivity' peak at zero delay) and is subsequently reflected at the open termination of the transmission line, producing a peak 5.5 ps after the directivity. The directivity and reflected signal have the same polarity, correctly indicating that the transmission line termination behaves as an open circuit. Additional structure, including a weaker peak at 11 ps, results from Fabry-Perot reflections between the open circuit and A-switch. The spectral amplitude and phase of both the directivity pulse D($\omega$) and the sample-reflected pulse R($\omega$) are isolated using appropriate Hann windows and computed by Fourier transform. We define usable bandwidth as frequencies for which the standard deviation is less than 5\% of the signal amplitude as shown in Fig. \ref{fig:signals}c, yielding a bandwidth of about 1 THz.

Characterization data for a sample board containing a bare transmission line (thru) are shown Fig. \ref{fig:signals}d-f. Accurate alignment of the transmission line allows the strength of the THz pulse reflected from the sample board interface (around 5.5ps) to be smaller than $<10$\% of the directivity, indicating good coupling between switch and sample boards. The transmitted THz pulse propagates across the sample board transmission line until the second interface, where it undergoes another partial reflection. This reflected pulse is observed at the A-switch about 42 ps after the directivity. The reflected signal is featureless between $\sim$20 and $\sim$42 ps. This window corresponds to a significant fraction of the transmission line length on the sample board, and sets the bounds for placement of the device to be tested.   

For both configurations, $D(\omega)$ is reproducible to within 5\% out to $\approx$1 THz (Fig. \ref{fig:signals}c and f). Recording $D(\omega)$ and $R(\omega)$ from the sample board interface (in addition to reflections from the device to be tested) allows us to characterize the emitted THz pulse and the sample board alignment. We consequently focus on THz reflection rather than transmission in what follows, even though our circuit permits the measurement of both THz reflection and transmission.

Having characterized the on-chip THz spectrometer at room temperature, we now demonstrate its cryogenic capability by extracting the complex optical conductivity ($\sigma(\omega)$) of a superconductor as it is cooled below its transition temperature. 
Conventional superconductivity exhibits distinct features such as the superconducting gap ($2\Delta$) in the real optical conductivity ($\sigma_1(\omega)$) and $1/\omega$ like response in the imaginary optical conductivity ($\sigma_2(\omega)$) \cite{Zimmermann1991}. 
To inform the design of the sample, we model a sample on the transmission lines as complex admittance connecting one metal line to the other. The admittance takes the form Y($\omega$) = Y$_1$($\omega$) + jY$_2$($\omega$) for radial frequency $\omega$ and imaginary unit $j$.The reflected signal R(YZ$_0$,$\omega$) is proportional to the directivity D($\omega$) and sample reflection coefficient $\Gamma(YZ_0, \omega)$, which is given by:

\begin{eqnarray}{\label{reflectCoeff}}
\Gamma(YZ_0, \omega) = \frac{- Y(\omega) Z_0}{2 + Y(\omega) Z_0} 
\end{eqnarray}

\noindent where $Z_0$ is the characteristic impedance of the transmission line.  
We are interested in maximizing the response to small changes in sample admittance $\delta YZ_0$, which can be expressed as $\delta\Gamma = \delta YZ_0 ({\partial\Gamma}/{\partial YZ_0})$. Using \eqref{reflectCoeff} for $\Gamma$, we find that this quantity is maximized when YZ$_0$ = 2. The lateral width and thickness of the superconducting film are thus chosen to satisfy the maximal contrast condition at temperatures just above the superconducting transition.

\begin{figure*}
\includegraphics[width=\textwidth]{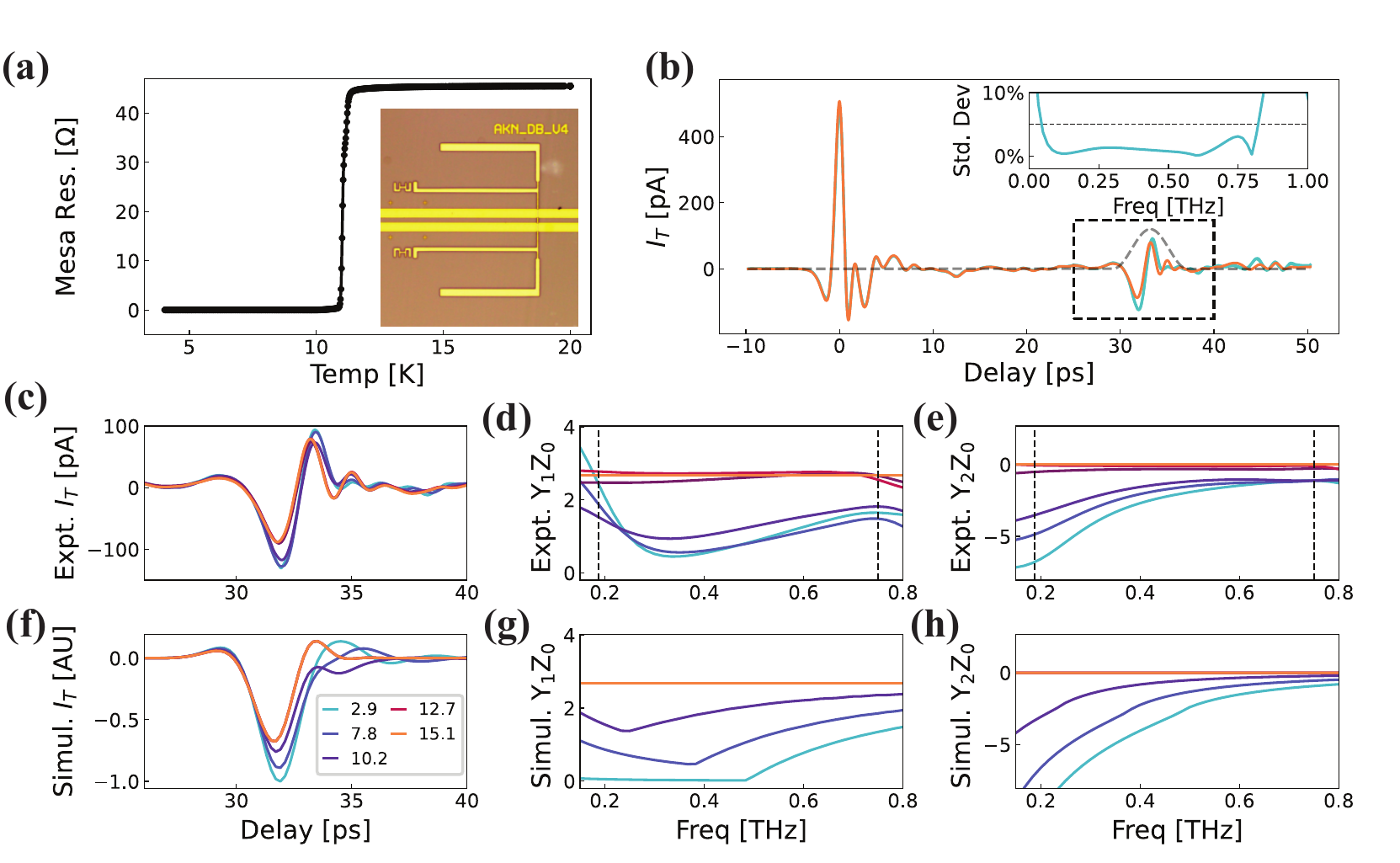}
\caption{On-chip THz spectroscopy of superconducting NbN. (a) DC, four-terminal measurement of 7.5 $\mu$m NbN mesa with inset showing sputtered NbN device on a DB (b) On-chip THz TDS of NbN mesa, with dashed lines drawn around NbN reflection (inset: representative 5\% magnitude of 810 GHz). Red and orange colors indicate temperatures above $T_c$; purple and blue colors indicate temperatures below $T_c$. (c) Windowed time-domain NbN reflection (d) experimentally-measured real admittance and (e) experimentally measured imaginary admittance (f) Zimmerman-simulated time domain transient (g) Zimmerman-simulated real admittance and (h) Zimmerman-simulated imaginary admittance.}
\label{fig:nbn}
\end{figure*}

For the superconducting film, we use NbN, chosen for its moderate superconducting transition temperature (T$_C$) and well-documented electrodynamics\cite{Cheng2016, Hong2013, Sim2017, Nuss1991, Nuss1998}. NbN films were deposited by magnetron sputtering (AJA ATC 2200-V sputtering system) with power 300W, pressure 1.8 mTorr, and 45/3.8 sccm Ar/N$_2$ gas flow. The four-terminal DC transport curve is shown in \ref{fig:nbn}a, with the data scaled by a geometric factor to accurately represent the resistance shorting the transmission line.
The sample has a T$_C$ of 11.1K, where the T$_C$ is defined as 90\% of the normal state resistance at 15K. With lateral width 7.5 $\mu$m, the sample is 3.2\% of one wavelength at 750 GHz, justifying the lumped element approximation. The sample thickness of 151 nm is chosen to bring the resistance in the normal state as close as possible to the maximum contrast condition. Experimentally, we find that the sample presented here has Y$Z_0$ = 2.58 at room temperature, which, given the residual resistivity ratio of 0.94 for NbN, ensures good matching at low temperatures. The sample board containing the NbN film is aligned to the switch board and cooled down to $T\approx$~3K in an exchange gas environment. The exchange gas ensures thermalization of the sample, despite the poor thermal conductivity of the COP switch board.

To extract the complex admittance of the superconducting state, we assume a frequency-independent admittance in the normal state, $Y_N$, which we take to be equivalent to the conductance measured by DC transport. The ratio of the reflected signal in the normal and superconducting states can then be used to extract the superconductor's admittance Y$_S$($\omega$) via the relation:

\begin{eqnarray}
Y_S(\omega)Z_0 = \frac{-2}{ 1 - \frac{R_N(\omega)}{R_S(\omega)}\frac{D_S(\omega)}{D_N(\omega) } \frac{2+Y_N Z_0}{Y_N Z_0} }
\label{eq:ys}.
\end{eqnarray}

The THz time-domain reflection signal of the NbN sample measured above and below the T$_C$ is presented in Fig. \ref{fig:nbn}b, with the dashed box outline highlighting the signal reflected from the NbN device.  The repeatability of the THz signal reflected from the sample at a temperature of 3K is shown in the inset, showing a usable bandwidth extending to $f\approx 750$ GHz. The dynamic range, defined as the ratio of the peak Fourier amplitude and the background noise level exceeds 2000 for the NbN reflection. 

Fig. \ref{fig:nbn}c shows the time-domain reflection from the NbN, after application of the window shown in gray in Fig. \ref{fig:nbn}b.  The measured reflection is temperature independent at high temperatures but exhibits strong temperature dependence after the onset of superconductivity; qualitatively, the peak signal increases while simultaneously also shifting to later times by $\delta t\leq 200$ fs as the temperature is lowered below $T_C$.  These behaviors are consistent with expectations. Specifically, the increase in the peak reflection signal results from decreased THz absorption by normal carriers, while the lagging phase arises from the inductive response of the incipient Cooper pairs\cite{Tinkham2004}.

The real ($Y_1(\omega)$) and imaginary ($Y_2(\omega)$) admittance, expressed in units of transmission line impedance ($Z_0$), are extracted using Eq. \ref{eq:ys} and are shown in Fig. \ref{fig:nbn}d and e, respectively. Free space THz conductivity measurements of NbN in the normal state report a featureless real conductivity equal to the measured DC value and a negligible imaginary conductivity \cite{Cheng2016, Sim2017} from DC to 750 GHz. We therefore use the DC value of Y$_N$ for all measurable frequencies.  

The upper and the lower frequency cutoffs are delineated with black dashed lines in Fig. \ref{fig:nbn}d and e. The upper frequency cutoff is defined as the frequency at which the normalized standard deviation of the THz signal magnitude is at least 5\% (inset), while the lower frequency cutoff is defined as the product of the minimum resolvable frequency of the windowed directivity or windowed NbN reflection and the equivalent noise bandwidth (ENBW) of the isolating Hann window of 1.5. The minimum resolvable frequency is given by the maximum size of the isolating window used for the reflection, which is set by the distance between the signal of interest and features arising from unwanted reflections such as those that occur at the switch board/sample board interface. Practically, in our current geometry the low frequency limit is 200 GHz. This could be be extended by increasing the physical spacing between the detector switch A and the switch-sample board interface.

We simulate the electromagnetic response of the superconductor using a model developed for disordered superconductors by Zimmerman and collaborators\cite{Zimmermann1991}.  We take the NbN to be in impure limit ($y=500$), use $T_C = 11.1$K, and use a superconducting gap $2\Delta(T)$ of the following form:

\begin{eqnarray}
2\Delta(T) = 2\Delta(0) \tanh{\left(\frac{\pi k_B T_c}{\Delta(0)}\sqrt{T_c/T - 1}\right)}
\label{eq:sc_gap}.
\end{eqnarray}

\noindent where, $2\Delta(0)$ is 500 GHz and $k_B$ is the Boltzmann constant. 
%The superconductor's simulated admittance was extended to negative frequencies by enforcing conjugate symmetry. 
The simulated time-domain response of the superconductor and the corresponding real and imaginary admittances are plotted in Fig. \ref{fig:nbn}f, g, and h, respectively. We use the complex admittance obtained from the BCS model\cite{Zimmermann1991} to simulate the reflected signal in the frequency domain, as shown in Fig. \ref{fig:nbn}g and h. The time domain signal shown in Fig. \ref{fig:nbn}f was then  reconstructed from the complex admittance using the inverse Fourier transform. 

The measured and the simulated time-domain profiles show similar qualitative features such as the increase in signal peak value and the phase lag below the superconducting transition temperature. The real admittance ($\propto \sigma_1(\omega)$) of the superconducting state reduces as we go from the high frequency ($\omega > 2\Delta$) response of the normal carriers into the gap ($\omega < 2\Delta$), followed by an increase due to thermal broadening of the superfluid peak at $\omega = 0$. The imaginary admittance ($\propto -\sigma_2(\omega)$) of the superconducting state shows the characteristic $ 1/\omega$-like response indicative of the superfluid density, whose spectral weight increases as the temperature decreases. 

While our results demonstrate the qualitative electrodynamic response of a superconducting thin film, further improvements in phase accuracy and bandwidth are needed to get a better quantitative understanding of, say, unconventional superconductors. Imperfect alignment of the laser spot and the PC switches, for instance, introduces a random variation in the THz emission and the propagation length of the THz pulses. We use the directivity ratio ${D_S(\omega)/D_N(\omega) }$ in Eq. \ref{eq:ys} to calibrate the system, however, the THz path error of length $\delta L$ subsequently accumulates to a phase error of $\omega \delta L/c$. Better phase sensitivity can likely be achieved by miniaturizing the photoconductive switches to match the laser spot. 

In summary, we have demonstrated a cryogenic-compatible on-chip THz spectrometer with a fast sample interchange architecture, and used it to extract the complex optical admittance of a deeply sub-wavelength superconducting NbN sample between 200 and 750 GHz. The confinement of the THz field within the transmission line and ease of sample interchange makes this spectrometer architecture ideal to investigate gate-tunable van der Waals heterostructures, with near-term applications to the physics of correlated electronic states and ultra-high frequency devices.

\begin{acknowledgments}
The authors acknowledge discussions with V. Ramaprasad, L.A. Cohen, and B. Potts, and the assistance of P. Kissin in designing the optical set up.   
The early stages of this work (through mid-2022) were supported by the Army Research Office under No. MURI W911NF-16-1-0361.  Work from 2022 until publication was supported by the National Science Foundation (NSF) Materials Research Science and Engineering Center (MRSEC) at UC Santa Barbara, Award DMR 1720256, through a seed grant and through the use of the MRL Shared experimental facilities.
This work was additionally supported by the Gordon and Betty Moore Foundation EPIQS program under award GBMF9471. 
A portion of this work was performed in the UCSB Nanofabrication Facility, an open access laboratory.  
A portion of this work was performed at the Institute for Terahertz Science and Technology (ITST) at UCSB. 
\end{acknowledgments}

\newpage 
\clearpage
\appendix
\section{COP and unbound substrate modes}

The generation of unbound leaky wave modes\cite{Shigesawa1995} greatly affects the ability of an on-chip THz spectrometer to measure the reflection and transmission from the device. Ideally, the THz pulse would propagate in only the CPS's bound quasi-TEM mode \cite{Gallagher2019}. If any coplanar transmission line and substrate dielectric environment supports leaky-wave modes, such modes are radiated isotropically upon emitter excitation. Some radiation goes into free space, some radiation is coupled to the transmission line as a bound mode, and some radiation is coupled as a leaky-wave mode in the substrate. All three types of modes are propagating THz electric fields that can be sampled by readout PC switches\cite{Lauck2022}. Unbound free space modes will propagate to the readout PC switch many picoseconds before the bound transient. The bound transmission line mode, then the unbound substrate modes, will reach the readout PC switch. Unbound substrate modes may propagate longitudinally in the substrate or may reflect from the substrate underside and edges, delaying their arrival at the readout PC switch by up to hundreds of picoseconds. As the substrate thickness is reduced, the substrate begins acting as a dielectric waveguide for unbound modes, producing measurable electric fields at long time delays. 

\begin{figure}
\includegraphics[width=\linewidth]{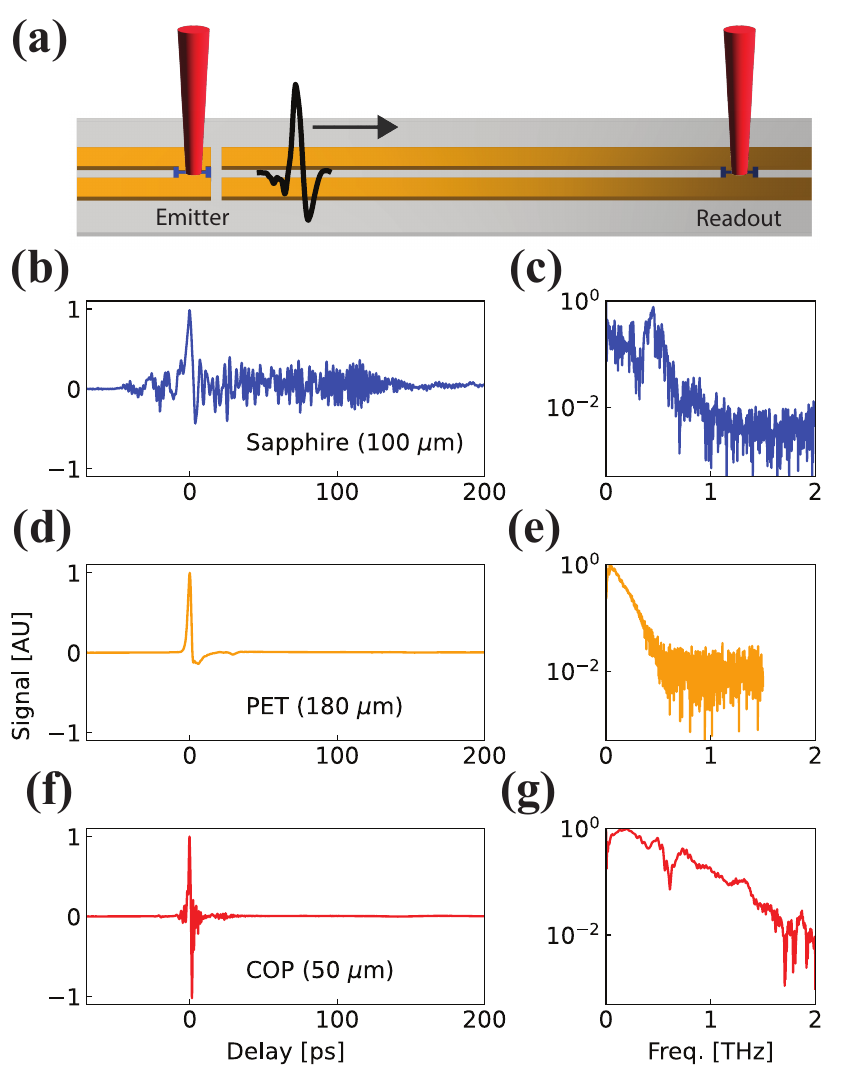}
\caption{Presence of unbound leaky-wave modes in alternate substrates. (a) Illustration of measurement schematic, showing a bound quasi-TEM transient propagating from the emitter to the readout, and time domain/frequency domain THz transients measured after 10 mm propagation with substrates: (b/c) 100 $\mu$m thick sapphire, (d/e) 180 $\mu$m thick PET and (f/g) 50 $\mu$m thick COP.}
\label{fig:leakyWaves}
\end{figure}

To demonstrate the effects of unbound leaky-wave modes, a thru circuit was fabricated with two LT GaAs switches placed 10 mm apart (Fig. \ref{fig:leakyWaves}a). The emitter is connected to the thru CPS of the same dimensions used in the main text via an capacitive coupler. The circuit was fabricated on three substrates: 50 $\mu$m thick COP, 180 $\mu$m thick polyethylene terephthalate (PET) and 100 $\mu$m thick sapphire. The resulting transients and their normalized Fourier transforms are shown in Fig. \ref{fig:leakyWaves}b-g.

The measured transmission with a sapphire substrate shows a signal 35 ps before the main transient from unbound free space modes. The bound transient, mixed with unbound modes, appears at zero delay. Strong oscillations continue for 130 ps, due to the leaky wave substrate modes and their reflections from the substrate underside and periphery. After 130 ps, the oscillations reduce in amplitude, but do not vanish. The bound transient cannot be isolated with a narrow window because that peak contains bound and unbound contributions. Even if the bound transient could be isolated, the window would have to be so narrow that the minimum resolvable frequency would be reduced to about 1 THz. 

The unbound mode cannot propagate for thin PET and COP substrates as a result of decreased substrate permittivity. The moderate loss tangent and dispersive index in PET, however, results in a bound transient with only 300 GHz bandwidth. COP's frequency-independent index and minimal loss tangent result in a large signal, large bandwidth transient, even after 10 mm of propagation. COP is therefore the best of the three thin substrates for THz applications.

In the case of short ($<1$ mm) transmission lines fabricated on thick ($>1$ mm) substrates, the reflected leaky-wave modes do not have sufficient time to propagate to the readout PC switch and are thus not conflated with the desirable bound mode signal. Both short transmission lines and thick substrates are undesirable. Short transmission lines increase the minimum resolvable frequency; thick substrates suffer exacerbated Cerenkov radiation loss. The use of COP substrates mitigates both factors, although at a cost in difficulty of switch fabrication.

\clearpage
\bibliography{refs}% Produces the bibliography via BibTeX.

\end{document}